\documentstyle[12pt,epsf]{article}
\newcommand{\be}{\begin{equation}}
\newcommand{\ee}{\end{equation}}
\newcommand{\bea}{\begin{eqnarray}}
\newcommand{\eea}{\end{eqnarray}}
\begin{document}
\begin{titlepage}
\hspace{12cm}KANAZAWA-97-20\\
\vspace{15mm}
\begin{center}
{\Large \bf Rapidly Converging Truncation Scheme of\\
the Exact Renormalization Group}
\end{center}
\vspace{0.5cm}
\begin{center}
Ken-Ichi AOKI \footnote{e-mail address: aoki@hep.s.kanazawa-u.ac.jp},
Keiichi MORIKAWA, Wataru SOUMA,\\ Jun-Ichi SUMI and Haruhiko TERAO
\footnote{e-mail address: terao@hep.s.kanazawa-u.ac.jp}
\end{center}
\begin{center}{\it Department of Physics, Kanazawa University\\
Kanazawa 920-1192, Japan}
\end{center}
\vspace{0.5cm}
\begin{center}{March~~1998}\end{center}
\vspace{30mm}
\begin{center} {\large \bf Abstract} \end{center}
\noindent
The truncation scheme dependence of the exact renormalization group
equations is investigated for scalar field theories in three dimensions.
The exponents are numerically estimated to the next-to-leading order of
the derivative expansion. It is found that the convergence property in
various truncations in the number of powers of the fields is
remarkably improved if the expansion is made around the minimum of the
effective potential. It is also shown that this truncation scheme is
suitable for evaluation of infrared effective potentials. The physical
interpretation of this improvement is discussed by considering $O(N)$
symmetric scalar theories in the large $N$ limit.
\end{titlepage}

\section{Introduction}
It has been well known for more than two decades that the Wilson 
renormalization group (RG) offers practical tools as well as profound
insights for investigation of non-perturbative phenomena in field
theories. The continuum versions of the Wilson RG equation are called
the exact renormalization group (ERG) equations, which are given in
the form of non-linear functional differential equations.
There have been proposed several formulations of the ERG, which are
found to be mutually equivalent. These ERG equations give the change
of the so-called Wilsonian effective actions\cite{WH,WK,P} ~ or 1PI
cutoff effective actions\cite{Wet1993,Bonini} ~ under scale variation
leaving the low energy physics unaltered. The Wilsonian effective
action may be regarded as a point in the infinite dimensional space of
theories, or the space of coupling constants, and the ERG generates
flows of the coupling constants in this space.

However, in practical use, it is inevitable to approximate such an
infinite dimensional theory space by a much smaller subspace in order
to solve the ERG equations. Needless to say, the non-perturbative
nature of the ERG should be maintained in this approximation. The
method generally applied is the so-called derivative expansion, which
expands the interactions in powers of the derivatives and truncates
the series at a certain order.\cite{Wetderi,Wet1,Morrisderi,Ball} ~ With
this approximation the full equation is reduced to coupled partial
differential equations. Recently, the ERG in the derivative expansion
approximation for scalar field theories has been extensively studied
at the order of the derivative squared and has been found to offer
fairly good non-perturbative results even
quantitatively.\cite{Berges,Morris1,Morris2,Morris3} ~ The lowest
order of the derivative expansion, neglecting all corrections to
derivative operators, is called the ``local potential
approximation'' (LPA).\cite{NCS,HH} ~ Although the wave function
renormalization and,
therefore, the anomalous dimension is ignored in the LPA, the leading
exponent $\nu$ is known to be estimated rather well by using the ERG.
However, the number of the couplings to be incorporated, or the number
of the beta functions, is still infinite in the derivative
expansion. If we consider application to more complicated systems
it would be favorable to approximate it further by a finite number of
couplings, as long as this is sufficiently effective.
This method is advantageous not only because it simplifies the
analysis but also because the effective couplings of physical interest
are treated directly. Such an approximation is naively performed by
truncated expansion of the Wilsonian effective action, in turn, in
powers of the fields. Actually it is found to work well practically
without loss of the non-perturbative nature as long as we adopt a
special expansion scheme, as is discussed later.

One of the advantageous points of the ERG is certainly that it is able
to allow for the systematic improvement of the approximations, as
mentioned above. However, the improvement totally relies on the
convergence of the results in non-perturbative analysis. We may
approve the results obtained by the ERG only after confirming their
sufficient convergence, since there is no small parameter which
controls the approximation. It should be noted here that the commonly
used expansion methods, e.g., the  perturbation theory, $1/N$
expansion and $\varepsilon$-expansion etc., generally produce asymptotic
series at best, in contrast to the convergence property exhibited by
the ERG.\cite{AMSSTeffective}

The main subject of this paper is the convergence of the expansion
scheme in terms of the fields. It has been claimed that this
convergence is rather poor,\cite{MOP} ~ or that the results cease to
converge.\cite{Morristrunc} ~ If such behavior appears commonly, it
would be a fatal defect of the ERG approach.\footnote{
The convergence of the derivative expansion in the
non-perturbative calculation remains difficult to see due to
complication in the higher orders. Morris examined this problem
perturbatively at two loops and found that the ERG with certain
cutoff profiles indeed displays convergence.\cite{Morris1}}
However, it has been realized that the expansion around the potential
minimum drastically improves this convergence
property.\cite{Wetderi,tetwet,Wet1,A} ~
Indeed, it is good news for the ERG approach that we may obtain
good convergence by adopting the appropriate truncation scheme.
However, it has not yet been investigated in detail how effective this
scheme is generally, nor has it been determined the origin of this
improvement.

In this paper we discuss convergence properties in different
expansions schemes by examining $Z_2$ symmetric scalar theories in 
three dimensions. As the physical quantities, the critical exponents
and also the infrared effective potentials, or the renormalized
trajectories, are compared among the different schemes. It is found
that the convergence property is significantly improved in the new
truncation scheme. We will also discuss the physical reason of the
improvement by studying $O(N)$ symmetric scalar theories in the large
$N$ limit. From this observation it is speculated that good
convergence depends on how accurately the relevant operator is covered
within the truncated subspace.

\section{Exact renormalization group equations}
First let us briefly review the formulations of the ERG with which
we will analyze the scalar theories. The ERG equation widely
studied recently is given by considering renormalization of the
so-called ``cutoff effective action'',
$\Gamma_{\mbox{eff}}[\phi]$.\cite{Wet1993,Bonini} ~ The cutoff
effective action is the 1PI part of the Wilsonian effective
action, namely, the generating functional of the connected and amputated
cutoff Green functions. Therefore the ERG equation may be obtained by
the Legendre transformation of the Polchinski equation for the
Wilsonian effective action.

In this formulation the cutoff is performed by introducing a proper
smooth function in terms of the momentum $q$ and the cutoff scale
$\Lambda$; $C(q,\Lambda)$, into the partition function as
\be
\exp W[J] = \int {\cal D}\phi \exp
\left\{ 
-\frac{1}{2} \phi \cdot C^{-1} \cdot \phi -S[\phi] + J \cdot \phi
\right\},
\label{partitionfunction}
\ee
where dot $(\cdot)$ denotes matrix contraction in momentum space.
From Eq.~(\ref{partitionfunction}) we may obtain the variation of $W$
with respect to the cutoff as \cite{P}
\be
\frac{\partial W[J]}{\partial \Lambda} =
-\frac{1}{2} 
\left\{
\frac{\delta W[J]}{\delta J}\cdot
\frac{\partial C^{-1}}{\partial \Lambda}\cdot
\frac{\delta W[J]}{\delta J}
+\mbox{tr}
\left(
\frac{\partial C^{-1}}{\partial \Lambda}\cdot
\frac{\delta^2 W[J]}{\delta J \delta J}
\right)
\right\}.
\ee
The ERG for the cutoff effective action $\Gamma_{\mbox{eff}}$ is
defined by the Legendre transformation:
\be
\Gamma_{\mbox{eff}}[\phi] + \frac{1}{2}\phi \cdot C^{-1} \cdot \phi
= -W[J]+J\cdot \phi.
\ee
By taking the canonical scaling under the shift of the cutoff scale
into consideration, the ERG equation for $\Gamma_{\mbox{eff}}$
in $D$ dimensions may
be written down as \cite{Morrisderi}
\be
\left(
\frac{\partial}{\partial t} 
+d_{\phi}\phi\cdot\frac{\delta}{\delta\phi}+\Delta_{\partial} -D
\right)
\Gamma_{\mbox{eff}}[\phi]
=
\frac{1}{2}  \mbox{tr}
\left\{
\frac{\partial C^{-1}}{\partial t} \cdot
\left(
C^{-1}+\frac{\delta^2 \Gamma_{\mbox{eff}}[\phi]}{\delta \phi \delta \phi}
\right) ^{-1}
\right\} ,
\label{legendrefloweq}
\ee
where $t =\ln ( \Lambda_0/\Lambda)$, and $d_{\phi}$ is the scaling
dimension of the scalar field which is given by $(D-2+\eta)/2$
with the anomalous dimension $\eta$.
The operator $\Delta_{\partial}$ counts
the number of the derivatives, which is given explicitly by
\be
\Delta_{\partial}
\equiv D + \int \frac{d^D q}{(2\pi)^D}
\phi(q)q_{\mu}\frac{\partial}{\partial q_{\mu}} 
\frac{\delta}{\delta\phi(q)}.
\ee

Thus the ERG equation is defined depending on the cutoff functions.
The physical quantities such as the exponents are found to be
independent of the cutoff scheme, as is expected.
However, this does not hold once some approximations have been
performed to these equations.\footnote{
In practice, as far as the scalar theories are concerned, this cutoff
scheme dependence is found to be rather weak, and the exponent changes
smoothly under variation of the cutoff
profiles.\cite{Ball,AMSSTcutoff}}
In this paper we are going to examine the extreme cases, i.e. schemes
with a very smooth cutoff and with the sharp cutoff limit.
We adopt the cutoff function
\be
C(q,\Lambda)=\frac{1}{q^2}\frac{\theta(q,\Lambda)}{1-\theta(q,\Lambda)},
\hspace{10mm}
\theta(q,\Lambda)=\frac{1}{1+(\Lambda^2/q^2)^2} \hspace{3mm}
\label{cutofffunction}
\ee
as the smooth one in the practical calculations,
{\it a la} Ref.~\cite{Morrisderi}.
The sharp cutoff version of the ERG will be discussed later.

In the derivative expansion,
the full ERG equation(\ref{legendrefloweq}) is reduced to a 
set of partial differential equations by substituting the effective 
action $\Gamma_{\mbox{eff}}$ of the form
\be
\Gamma_{\mbox{eff}}[\phi]=\int d^D x \left\{ 
V(\phi;t)+\frac{1}{2}K(\phi;t)(\partial_{\mu}\phi)^2 + \cdots~~~~
\right\},
\ee
where $V(\phi;t)$ and $K(\phi;t)$ are cutoff-dependent functions. In the
second order of the expansion we may take the variation of the first two 
terms only, ${\partial V}/{\partial t}$ and ${\partial K}/{\partial
t}$.
For the cutoff function given by (\ref{cutofffunction}),
these ERG equations in three dimensions are found to be
\bea
\frac{\partial V}{\partial t}
&=&-\frac{1}{2}(1+\eta)\phi V'+3V
-\frac{1-\eta/4}{\sqrt{K}\sqrt{V''+2\sqrt{K}}} \hspace{3mm}  ,
\label{dVdt} \\
\frac{\partial K}{\partial t}
&=&-\frac{1}{2}(1+\eta)\phi K'-\eta K
+(1- \frac{\eta}{4})(\frac{1}{48}
\frac{24KK''-19(K')^2}{K^{3/2}(V''+2\sqrt{K})^{3/2}}  \nonumber \\
& &-\frac{1}{48} \frac{58V'''K' \sqrt{K} +57(K')^2 +(V''')^2 K}
{K(V''+2 \sqrt{K})^{5/2}}   \nonumber \\
& &+\frac{5}{12} \frac{(V''')^2 K+2V'''K' \sqrt{K} + (K')^2}
{\sqrt{K} (V''+2 \sqrt{K})^{7/2}}
) \hspace{1mm}  ,
\label{dKdt}
\eea
where the prime denotes differentiation
with respect to $\phi$. \cite{Morrisderi} ~ The anomalous dimension
$\eta(t)$ is determined by imposing
the renormalization condition for the wave function, $K(\phi=0;t)=1$.
In the LPA we may solve only Eq.~(\ref{dVdt})
with respect to the effective potential $V(\phi;t)$
by reducing $K=1$ and, therefore, $\eta=0$.
Actually these partial differential equations have been solved and
found to give the exponents with fairly 
good accuracy.\cite{Morris1,Morris2,Morris3,Morristrunc} ~ In the next
section these equations are examined by expansion in powers of the
fields.

In the case of the sharp cutoff scheme, we examine the 
Wegner-Houghton (WH) equation\cite{WH} instead of
Eq.~(\ref{legendrefloweq}) with the sharp cutoff limit.
The WH equation is formulated so that the fields are integrated
gradually from the modes with higher momentum by introducing the sharp
cutoff into the path integral measure. Indeed, the ERG equation for the
1PI effective action as well as the  Polchinski equation\cite{P} turns
out to be equivalent to the WH equation in the sharp cutoff
limit.\cite{MorrisIJMP,AMSSTcutoff} ~
However, the ERG equation is known to exhibit non-analytic dependence
on the momentum in this limit. Therefore we examine the sharp cutoff
scheme only in the LPA. The WH equation in the LPA is given simply by
\be
\frac{\partial V}{\partial t}=
3V-\frac{1}{2}\phi V'+\frac{1}{4\pi^2} \ln(1+V'') \hspace{3mm}  ,
\label{LPAWH}
\ee
in three dimensions.\cite{NCS,HH}

These partial differential equations, of course, may be solved
directly. However, this would not be practical for more
complicated systems. Indeed, it turns out to be much more economical
to solve them by reducing them to coupled ordinary differential 
equations. Besides, if we are interested in the effective 
coupling constants of the theories, e.g. masses,
self-interactions, gauge couplings and so on, it is
natural to solve the ERG equations for these coupling constants
by expanding the effective action into a sum of operators.
Note that Equations (\ref{legendrefloweq}) and (\ref{LPAWH}) given
in the two different cutoff schemes do not produce the same results
for physical quantities even if both are analyzed in the LPA. 
Rather, in this paper we are interested in the convergence properties
of the solutions of the equations obtained by truncation of their
power expansion in each case. In the next section the convergence
behavior is explicitly examined.

\section{Truncation in the comoving frame}
First let us examine the ERG equation (\ref{LPAWH})
for the $Z_2$ symmetric scalar theory. If we approximate the 
effective potential $V(\phi ;t)$ by a finite order power series 
in terms of the $Z_2$ invariant variable $\rho=\phi^2/2$ (Scheme I),
\be
V(\rho; t)=\sum _{n=1}^{M} \frac{a_n(t)}{n!} \rho ^n,
\ee
then we obtain $M$ ordinary differential equations for the running
couplings $a_n$.\footnote{
The $M$ coupled beta functions lead us to $M$ distinct fixed point
solutions, all of which but two (the trivial fixed point and the
so-called Wilson-Fisher fixed point) should be fake in this
approximation.
However, we may easily identify the true fixed point among these
solutions by looking at their stability against the truncation.}
We may suppose naively that the results obtained with these truncated
equations converge to the solutions of the partial differential
equation (10) as the order of the truncation $M$ is increased. 
However, this is not the case.
In Fig.~1 the truncation dependence of the leading exponent $\nu$ is
shown. It is seen that the solutions cease to converge beyond a
certain order and finally to oscillate with 4-fold periodicity around
the expected value from the partial differential equation (10).
Actually, the non-trivial fixed point itself is also found to oscillate
similarly in the truncated approximation.
Morris pointed out in Ref.~\cite{Morristrunc}
that this oscillatory behavior is related to the singularity of the
fixed point solution of Eq.~(10) in the complex plane.
The singularities $\rho_*=|\rho_*|e^{i \theta_*}$ closest to the
origin are located at $(|\rho_{*}|,\theta_*)=(0.123,\pm 0.514 \pi)$.
The existence of these singularities implies that the coefficients of
the fixed point solution expanded in powers around the origin appear
with 4-fold oscillation and that the convergence radius of the
expansion series is given by 0.123.
From these observations Morris has explained the behavior of the
truncated solutions in Scheme I and that the leading exponent cannot
converge to a definite value with precision beyond an error of 0.008.
\begin{figure}[htb]
\epsfxsize=0.85\textwidth
\centerline{\epsffile{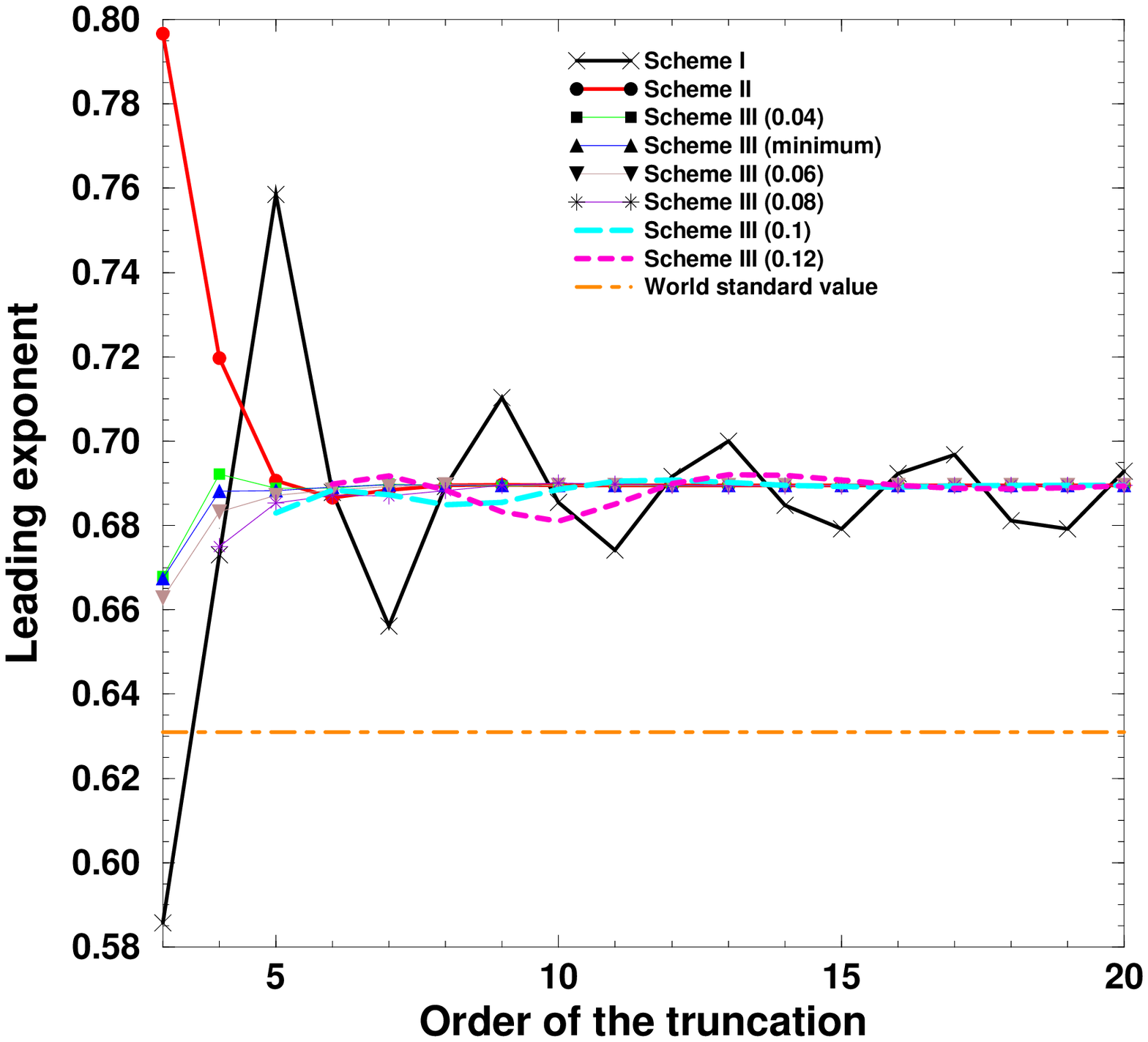}}
\vspace{3mm}
\centerline{
\parbox{140mm}{
\footnotesize 
Fig.~1~~The truncation dependence of the leading exponent evaluated
using various truncation schemes. The solutions not displayed
($M=3$ of Scheme III (0.08), $M=3,4$ of Scheme III (0.1) and
$M=3,4,5$ of Scheme III (0.12))
lie outside the range on the vertical axis.
The values in the parentheses in the legend correspond to the
expansion point $\rho_0$ in Scheme III. 
The term ``minimum'' means the minimum of the fixed point potential
in each truncation.
}
}
\end{figure}
\vspace{3mm}

This undesirable behavior, however, is drastically improved if we
expand the potential around its minimum. The effective potential may
be approximated in turn by (Scheme II)
\be
V(\rho; t)=\sum _{n=2}^{M} \frac{b_n(t)}{n!} (\rho - \rho _{0}(t))^n,
\ee
where $\rho_0$ is the potential minimum. Therefore here the term
linear in $\rho-\rho_0$ is absent.
Indeed, as is seen in Fig.~1 (see also Fig.~4), the results for the
exponent calculated in this scheme converge very rapidly to
0.689459056$\pm$2, which is consistent with the results obtained by
analysis of the partial differential equation (10), 0.689459056.
(The latter will be reported elsewhere.)
Thus we may say that the expansion scheme II is a quite effective
method in obtaining an accurate answer in a fairly small-dimensional
subspace. The fixed point potential obtained with this
method (Scheme II) is also shown in Fig.~2 (right).
It is rather surprising that the potential of the fixed point solution
can be obtained within a certain range of the field variable $\rho$
quite well with such a simple analysis.
\begin{figure}[htb]
\parbox{77mm}{
\epsfxsize=0.45\textwidth
\begin{center}
\leavevmode
\epsfbox{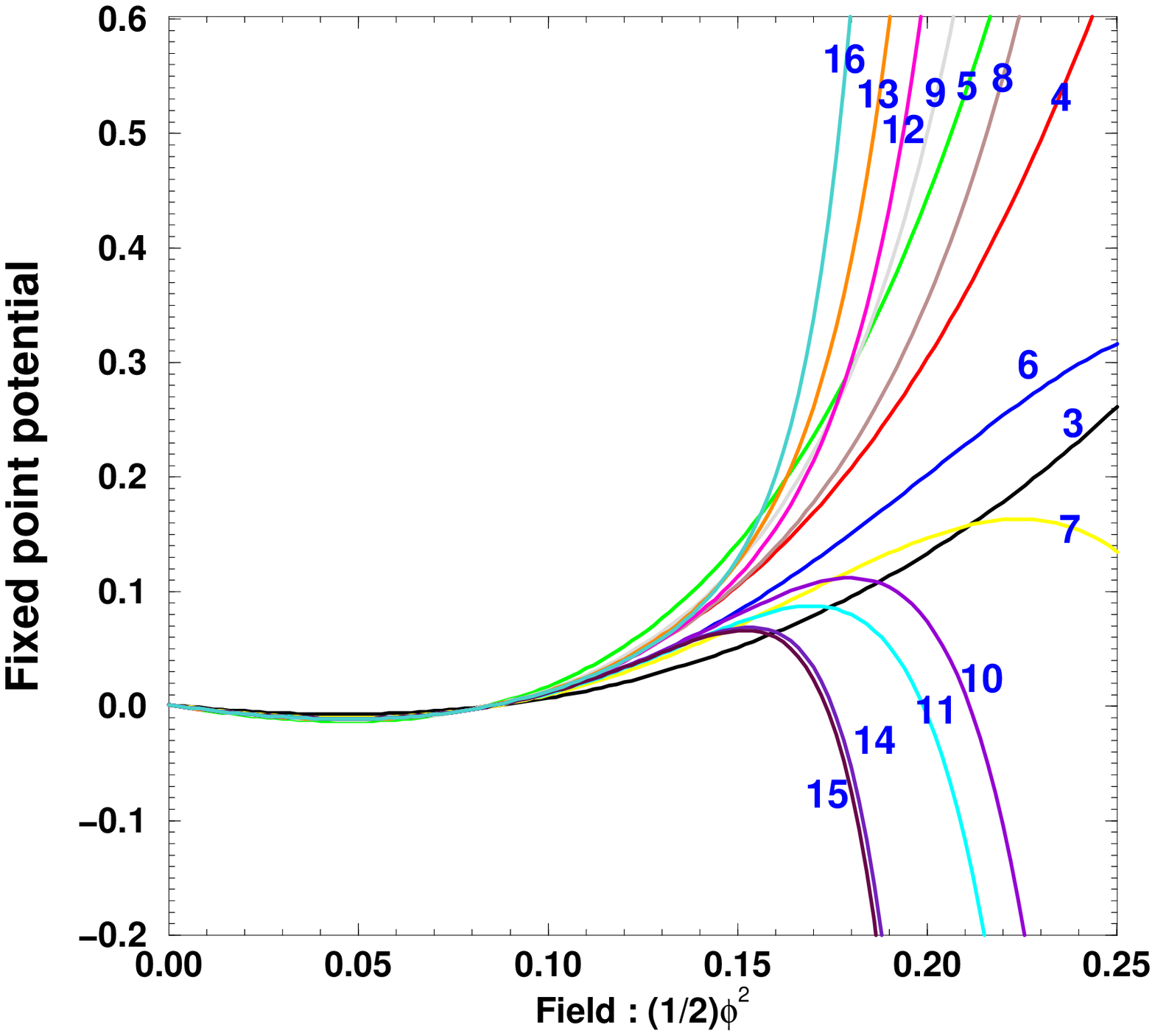}
\end{center}
}
\parbox{77mm}{
\epsfxsize=0.45\textwidth
\begin{center}
\leavevmode
\epsfbox{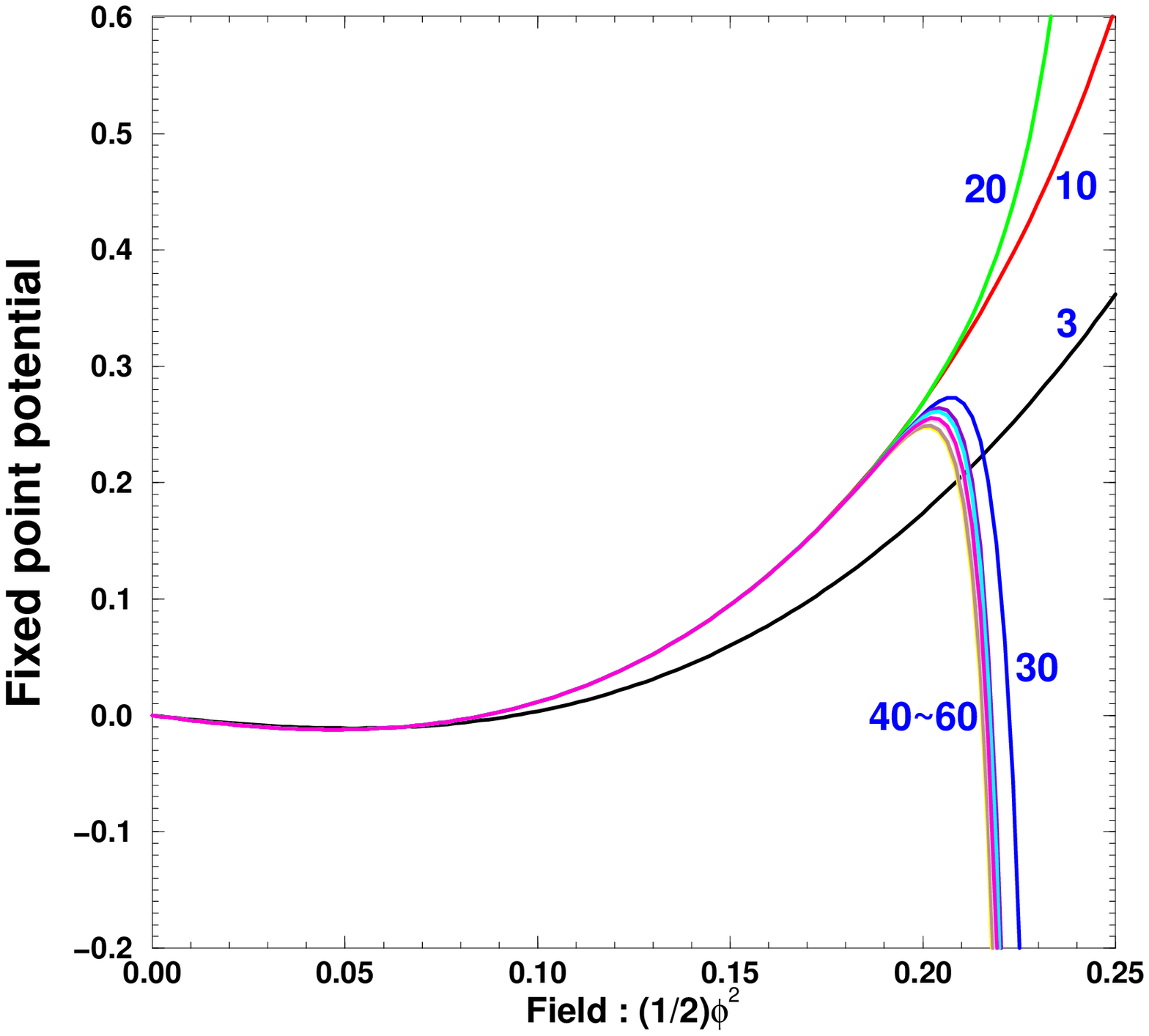}
\end{center}
}
\end{figure}
\vspace{-8mm}
\begin{center}
\parbox{140mm}{
\footnotesize
Fig.~2~~The truncation dependences of the fixed point potential
evaluated using Schemes I and II are shown to the left and right in
the figure, respectively. The integers in the figures denote the
orders of the truncation.
}
\end{center}
\vspace{5mm}

However, the series of the solution obtained in truncated approximation
of Scheme II does not converge perfectly. We examined the leading
exponent $\nu$ in this analysis up to the 60th order of truncation.
The logarithmic plot of the obtained coefficients $\rho _0$ and $b_n$
of the fixed point solution and the leading exponent against the order
of truncation is shown up to 60th in Figs.~3 and 4, respectively.
It is seen that the truncation dependence does not disappear
completely, and the results display oscillatory behavior with 3-fold
approximate periodicity, as is shown in Figs.~3 and 4.
These behavior can be clearly understood by considering the
singularities of the untruncated fixed point
solution.\cite{Morrisprivate}
Since the minimum of the fixed point potential is at
$\rho_0 =0.0471$, we have $|\rho_* - \rho_0|=0.134$ and
${\rm arg}(\rho_* - \rho_0)=0.64 \pi$.
This angle is close to $2\pi/3$, which tells us that
the truncated solution in Scheme II oscillates with 3-fold periodicity.
Also, the expansion should have a finite convergence radius of 0.134.
Therefore the boundary of the convergence radius appears around
$\rho =0.181$. This is indeed seen in Fig.~2 (right).
Although Scheme II extends the convergence radius slightly,
the convergence of the leading exponent is much improved up to the
error $10^{-9}$, as is seen in Fig.~4 (left).\footnote{
It is noted that in the analysis of Eq.~(10) the minimum expansion
in terms of the variable $\phi$ becomes worse than Scheme I
due to the smaller convergence radius.
}
\begin{figure}[htb]
\parbox{77mm}{
\epsfxsize=0.45\textwidth
\begin{center}
\leavevmode
\epsfbox{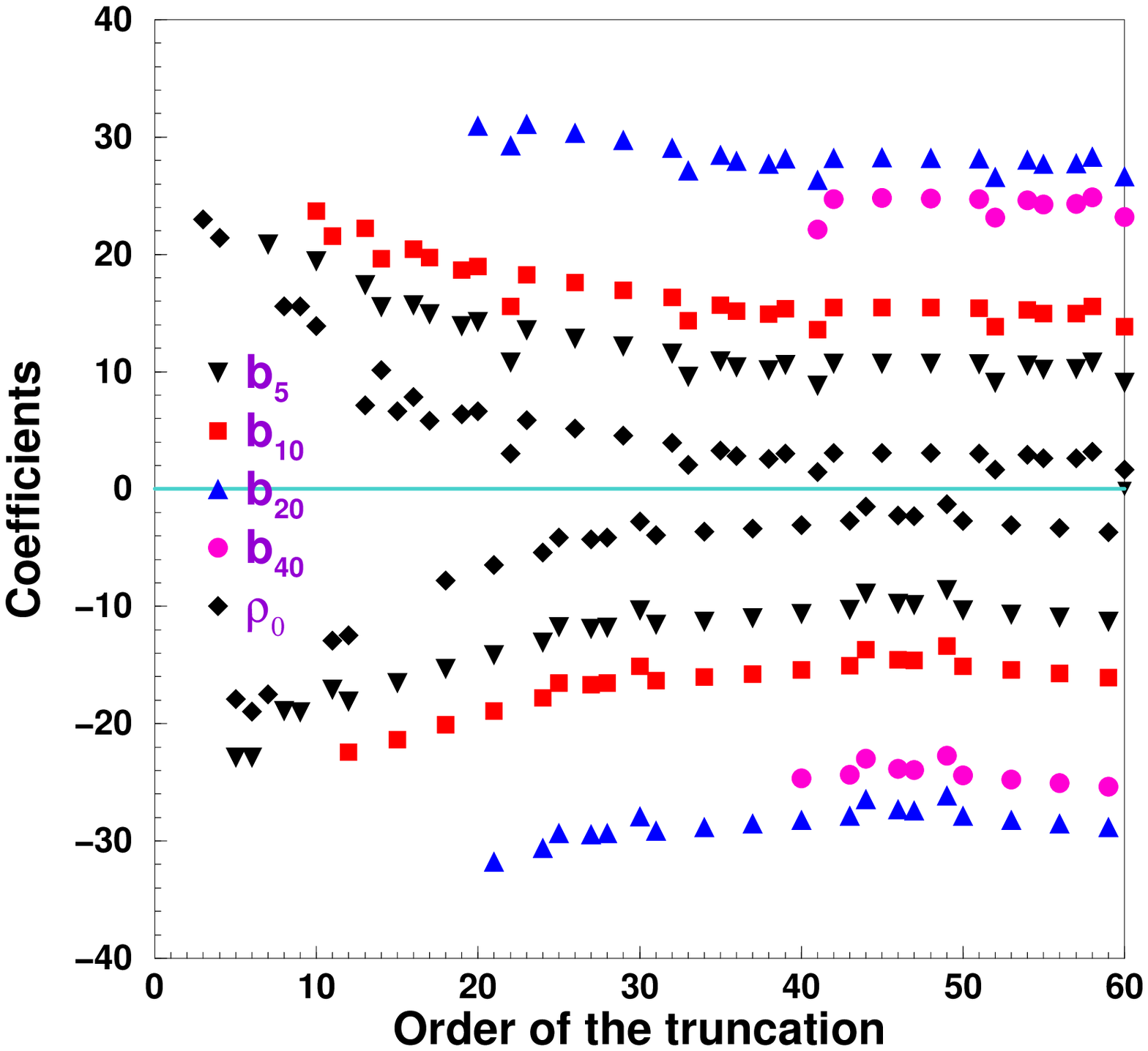}
\end{center}
}
\parbox{77mm}{
\epsfxsize=0.48\textwidth
\vspace{2mm}
\begin{center}
\leavevmode
\epsfbox{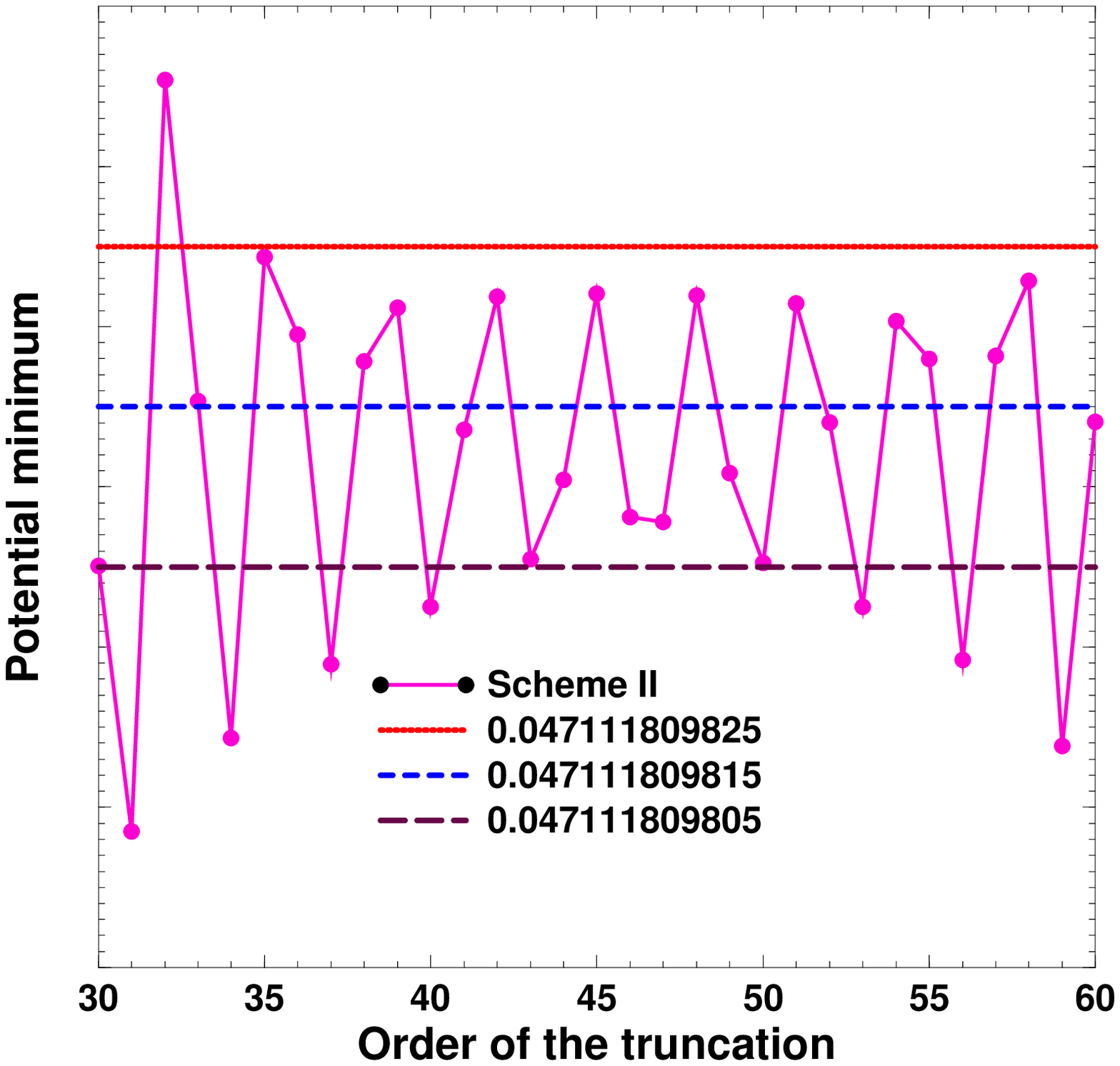}
\end{center}
}
\end{figure}
\vspace{-8mm}
\begin{center}
\parbox{140mm}{
\footnotesize 
Fig.~3~~The large order behavior of the coefficients $b_n$
and the potential minimum $\rho_0$ evaluated using Scheme II.
The vertical axis of the left figure denotes
$\ln (|b_i-\langle b_i \rangle |10^{11}+1)
{\rm sign}(b_i - \langle b_i \rangle)$.
}
\end{center}
\begin{figure}[htb]
\parbox{77mm}{
\epsfxsize=0.45\textwidth
\begin{center}
\leavevmode
\epsfbox{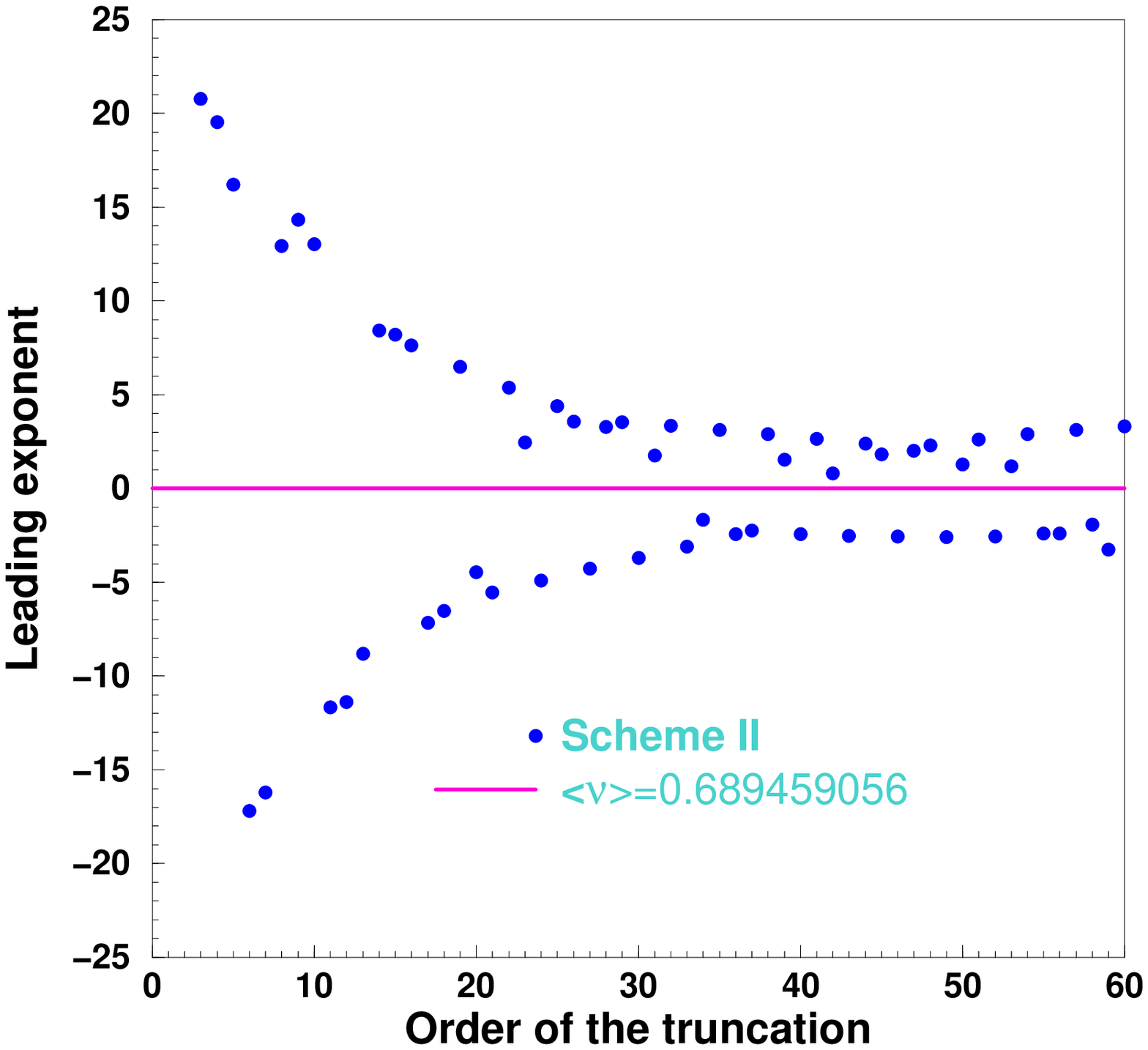}
\end{center}
}
\parbox{77mm}{
\epsfxsize=0.45\textwidth
\begin{center}
\leavevmode
\epsfbox{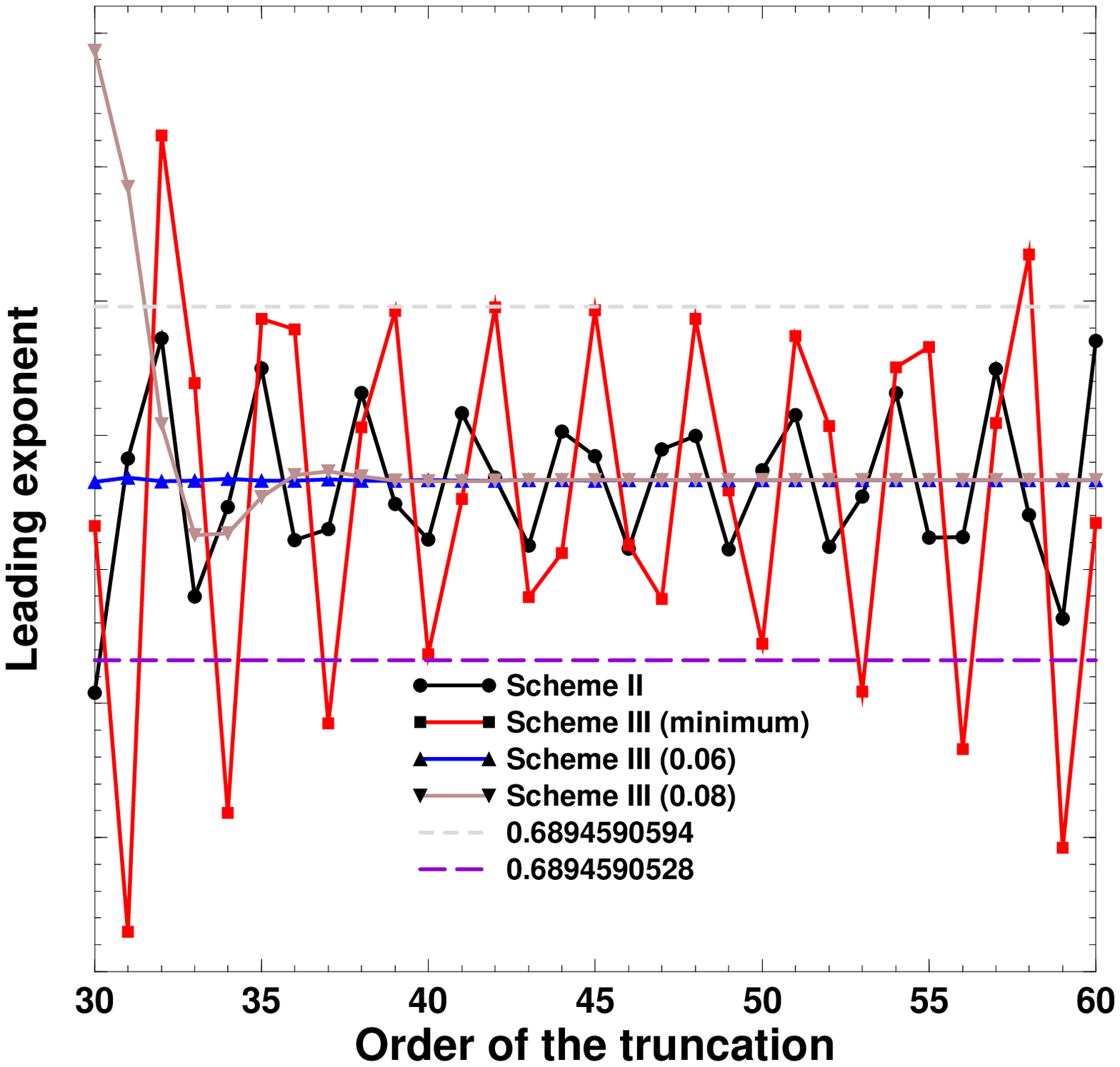}
\end{center}
}
\end{figure}
\vspace{-8mm}
\begin{center}
\parbox{140mm}{
\footnotesize 
Fig.~4~~The large order behavior of the leading exponent evaluated
using Scheme II and III. The vertical axis of the left figure denotes
$\ln( |\nu - \langle \nu \rangle|10^{10}+1)
{\rm sign}(\nu - \langle \nu \rangle)$.
}
\end{center}
\vspace{5mm}

Such truncation dependence in the two schemes is seen in the ERG
equations with smooth cutoff, (\ref{dVdt}) and (\ref{dKdt}), as well.
In Fig.~5 the leading exponent obtained by Scheme I and Scheme II in 
the LPA is shown. We find that Scheme II is again clearly 
superior to Scheme I, while the oscillation, even in Scheme I,
is significantly attenuated in the smooth cutoff scheme.
The value to which the leading exponent converges is 
0.660.\footnote{
The partial ERG equations given by (\ref{dVdt}) and (\ref{dKdt}) 
are examined in great detail in Ref.~\cite{Morris1,Morris2}.
Our results are consistent with those of that analysis.}
In the second order of the derivative expansion we examined the
truncation dependence of the leading exponent and also of the anomalous
dimension in Scheme II. 
\footnote{The exponents could not even be evaluated in Scheme I.} 
In this analysis the function $K(\rho;t)$ is also expanded around
$\rho_0$ and is truncated at the same order  as $V(\rho;t)$.
The results are shown in Figs.~5 and 6.
The values so obtained are $\nu=0.617476$ and $\eta=0.05425$.
The world standard values are $\nu=0.6310$ and $\eta=0.0375$ from the
$\varepsilon$-expansion.\cite{ZJ} ~ It is worth while to mention
that the leading exponent indeed approaches the world standard
value when going up to the next-to-leading order of the derivative
expansion.
\begin{figure}[htb]
\parbox{77mm}{
\epsfxsize=0.45\textwidth
\begin{center}
\leavevmode
\epsfbox{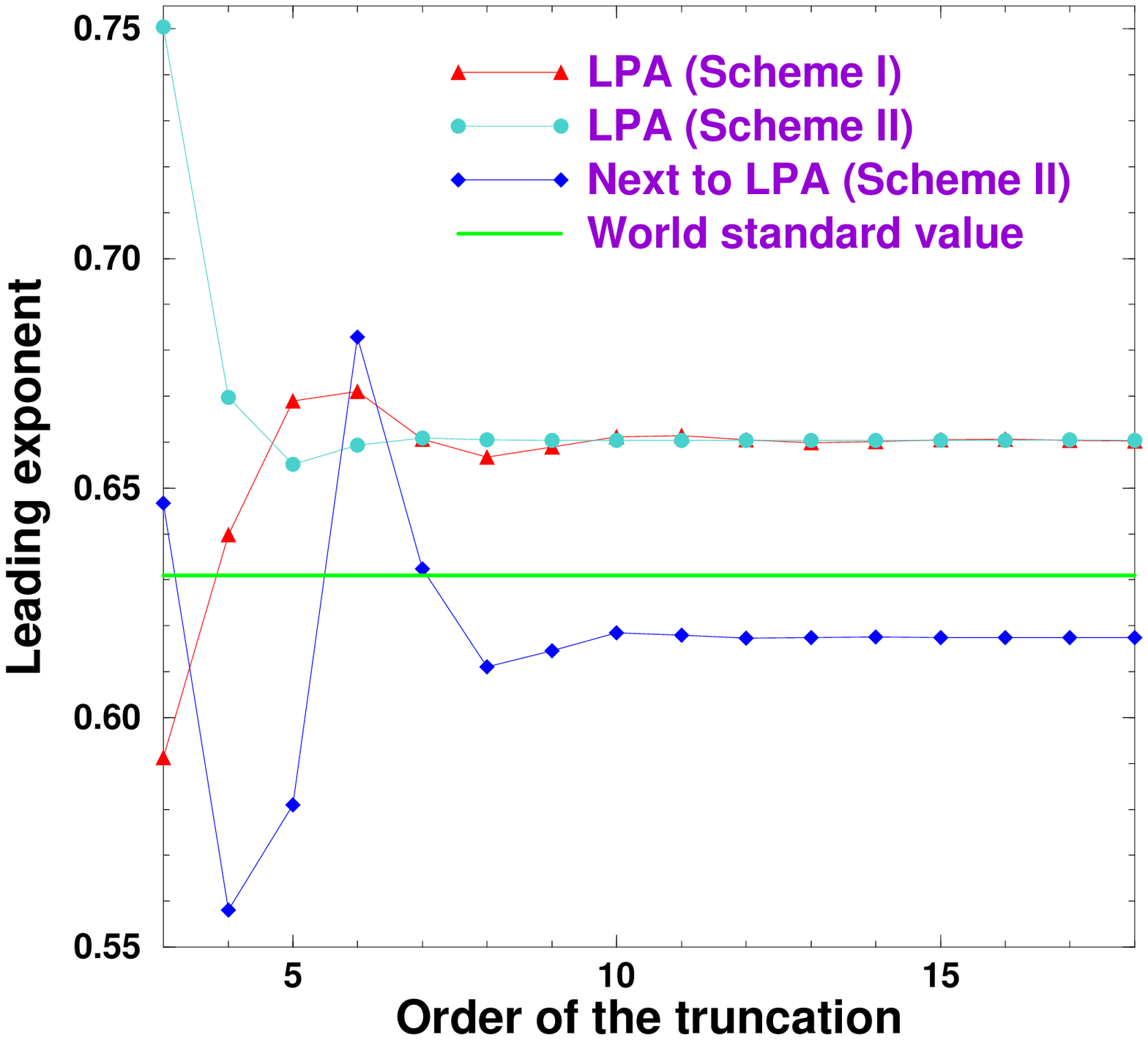}
\end{center}
\begin{center}
\parbox{63mm}{\footnotesize
Fig.~5~~The leading exponent evaluated using Scheme II}
\end{center}
}
\parbox{77mm}{
\epsfxsize=0.45\textwidth
\begin{center}
\leavevmode
\epsfbox{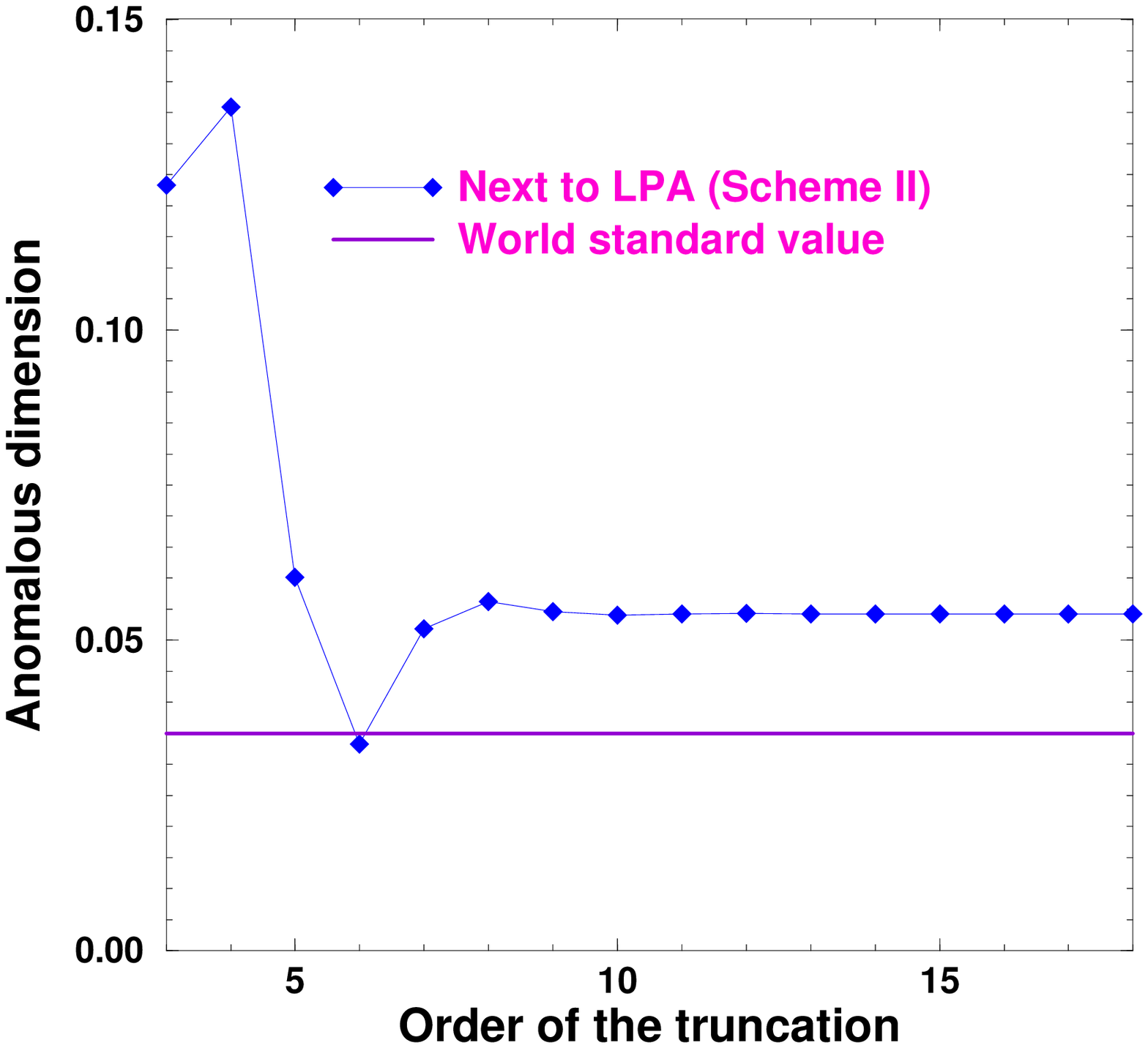}
\end{center}
\begin{center}
\parbox{65mm}{\footnotesize 
Fig.~6~~The anomalous dimension evaluated using Scheme II}
\end{center}
}
\end{figure}

In general $d/dt$ in the ERG defines a vector field of RG flow on the 
theory space and is given in the coordinate system $\{g_i\}$ by
\be
\frac{d}{dt}=\beta_i(g)\frac{\partial}{\partial g_i},
\ee
where we have introduced the generalized beta functions
$\beta_i(g)=dg_i/dt$. Correspondingly, the variation of the effective
action $S$ may be written as
\be
\frac{dS}{dt}=\beta_i(g)\xi_i(g),
\ee 
where $\xi_i(g)=\partial S/\partial g_i$ are the base vectors. In
Scheme II the base vectors are dependent on the position in the theory
space, while they are fixed in Scheme I. We refer to these
coordinate systems as the ``comoving frame'' and  ``fixed frame'',
respectively.\cite{AMSSTeffective}

Note that Schemes I and II with same maximum powers $M$ employ the
same subspace of the polynomials. We project the flow in the full
theory space on the subspace and evaluate the beta functions for the
projected flow. The projection depends on the choice of the
coordinates, or on the manner of expansion. In fact, the beta
functions in the truncation Schemes I and II are evaluated for
different projections. This causes deviation of the results between
the two schemes discussed here.

From the preceding argument it may be expected that more accurate
convergence is achieved if we expand at a point with a larger
convergence radius. In order to see this, we examine another type of
truncation, similar to Scheme I, but with the expansion point shifted
from the origin by some fixed values $\rho _0$ (Scheme III):
\be
V(\rho; t)=\sum _{n=1}^{M} \frac{c_n(t)}{n!} (\rho - \rho _{0})^n.
\ee
This scheme is an example of the fixed frame. The leading exponent
obtained in this approximation is also shown in Fig.~1 and in Fig.~4
(right). Here we employ Eq.~(\ref{LPAWH}). It is reasonable that the
value will converge to a definite value with high precision if the
expansion point $\rho_0$ is far from singularities. Indeed, we can
obtain a convergent value of leading exponent with precision on the
order of $10^{-16}$, 0.68945905616213484$\pm$1, by choosing
$\rho_0=0.08$. However, in order to obtain a highly accurate value, we
need a very large order of truncation.

One should employ a truncation method which provides a result with
sufficient convergence and precision even with a small order of
truncation. As is seen in the next section, in the large $N$ limit,
Scheme II is found to give us the exact solution. Apart from this
extreme case of large $N$ theory, however, it is significant for the
practical analysis of complicated systems to be converging with small
order of truncation, because it is in general difficult to evaluate at
the large order of truncation. It is seen from Fig.~1 that we should
set the value of $\rho_0$ to the minimum of the fixed point potential
if we demand better convergence at small order of truncation. 
If the highly accurate precision is not required, or if it is
difficult to analyze with a large order of truncation, the truncated
approximation of Scheme II is sufficiently workable owing to its
simplicity. The fact that Scheme III with $\rho_0$ set to the minimum
of the fixed point potential also gives good convergence in the small
subspace implies that the improvement of the approximation originates
in the choice of the base vectors $\xi_i$ around the fixed point.
This seems reasonable, since the exponent is determined solely by the
structure of the ERG equation around the non-trivial fixed point.
Needless to say, the truncated approximation of the comoving frame
is much more advantageous in practical analysis than that of the fixed
frame, since the position of the minimum of the fixed point potential
cannot be known {\it a priori}.

\section{Large $N$ limit}
If we extend the observation examined in the previous section to
$O(N)$ symmetric scalar theories in three dimensions, then the
approximation in Scheme II is found to show stronger convergence as
$N$ increases, while the exponents obtained in Scheme I become more
fluctuating.\cite{WH,MOP,Changpr,RTWlargen,Wet1,A,Lit,Comeon,Morlargen,Morris3}
~Moreover, the truncation dependence turns out to
disappear completely in the large $N$ limit, as discussed in
Ref.~\cite{AMSSTeffective}. Therefore we discuss the physical
reason behind this remarkable improvement of convergence by
considering $O(N)$ models in the large $N$ limit.

The LPA WH equation in $D$ dimensions is reduced in the large $N$
limit to 
\be
\frac{\partial V}{\partial t} = 
DV+(2-D)\rho V_{\rho}+ \frac{A_D}{2}\ln(1+V_{\rho}),
\label{largeNLPAWH}
\ee
where $\rho$ denotes $\Sigma_{a=1}^N(\phi^a)^2/2$, $V_{\rho}$ denotes
the differentiation of $V$ with respect to $\rho$ and $A_D$ is the
surface of the $D$-dimensional unit sphere divided by $(2\pi)^D$.
Here we have rescaled $\rho$ and $V$ properly by $N$ before taking the
limit. It is known that Eq.~(\ref{largeNLPAWH}) gives the exact
effective potential in the large $N$ limit. If we expand
Eq.~(\ref{largeNLPAWH}) in Scheme II, the resulting differential
equations,
\bea
& & \frac{d\rho_0}{dt}\equiv \beta_1 
=(D-2)\rho_0-\frac{A_D}{2}, \nonumber \\
& & \frac{db_2}{dt}\equiv \beta_2 
=(4-D)b_2-\frac{A_D}{2}b_2^2, \\
& & \frac{db_3}{dt}\equiv \beta_3 
=(6-2D)b_3-\frac{A_D}{2}(3b_2b_3-2b_2^3),
~~~ {\rm etc.}, \nonumber 
\eea
are found to be analytically soluble order by order.
This means that the entire flow diagram in the theory space can be
exactly determined within any finite dimensional subspaces.
We refer to such a special truncation scheme as the ``perfect
coordinates''.\cite{AMSSTeffective} Generally, it would be difficult
to find out such coordinates that enable us to solve the ERG equations
exactly. However, Scheme II turns out to be an example of perfect
coordinates in the large $N$ limit. Such a remarkable simplification
does not occur in Scheme I.

Due to the perfectness of the coordinates employed, the exact values
for the exponents can be obtained. The exponents are given by the
eigenvalues of the matrix $\Omega_{ij}=\partial \beta_i/\partial b_j$
evaluated at the non-trivial fixed point:
\be
\Omega=\left(
\begin{array}{cccc}
D-2    & 0                                  &  0     & \cdots \\ 
0      & D-4                                &  0     & \cdots \\ 
0      & \frac{24}{A_D} \frac{(4-D)^2}{6-D} & D-6    & \cdots \\ 
\cdots & \cdots                             & \cdots & \cdots
\end{array} 
\right).
\ee
Thus the exponents are exactly determined from the eigenvalues, 
or the diagonal elements of this matrix, and $\nu=1/(D-2)$. 
Also the corresponding eigenvectors of this matrix give us the
so-called relevant and the irrelevant operators.
The most characteristic feature of the ERG equations in Scheme II 
is that the relevant coupling precisely coincides with $\rho_0$ and is
not influenced by the truncation. We can say that this is the direct
reason why the leading exponent is calculated in a
truncation-independent way.

Thus the relevant operator corresponding to the coupling $\rho_0$ 
has been found to be given by $V_{\rho}^*$. In addition to the
relevant operator, all eigen-operators can be derived exactly from
Eq.~(\ref{largeNLPAWH}) as follows. Suppose $V^*(\rho)$ is the
non-trivial fixed point solution of the ERG equation and consider an
infinitesimal deviation from this; $V(\rho;t)=V^*(\rho)+\delta
V(\rho;t)$. Then we obtain the eigenvalue equation with respect to
$\delta V$ as
\be
D\delta V - 2\frac{V^*_{\rho}}{V^*_{\rho\rho}}\delta V_{\rho}
=\lambda \delta V.
\ee
By solving this equation, all of the eigenvectors are found to be 
given by
\be 
\delta V(\rho;t) \propto (V^*_{\rho}(\rho))^{\frac{D-\lambda}{2}}.
\ee
If we demand the analyticity of $\delta V$, then the eigenvalues
are determined to be $\lambda=D-2,D-4,D-6,\cdots$, as expected.

Moreover, it turns out to be possible to reformulate the ERG so that
the effective potential $V(\rho;t)$ is expanded into a power series of
the (ir)relevant operators $(V_{\rho})^n$. For this purpose, let us
first introduce two auxiliary fields $\chi$ and $\eta$ to the theory,
\be
Z= \int {\cal D}\phi^a {\cal D}\chi {\cal D}\eta
\exp \left\{
-\int d^D x \left[ \frac{1}{2}(\partial _{\mu}\phi^a)^2 
+ \chi\left( {1\over 2}\left(\phi^a \right)^2  - N \eta \right)
+N V(\eta)
\right]\right\}.
\ee
In the large $N$ limit, the path integral of these auxiliary fields is
evaluated by the saddle point method. Then the effective potential
$V({\bar \eta};t)$ is given by solving the coupled equations
\bea
V(\bar{\eta};t) &=&
\frac{1}{2} {\rm trln} (- \Box  + \chi)
+ \chi (\bar{\eta} - \eta) + V(\eta) ,\\
\eta &=&
\bar{\eta} + \frac{1}{2} {\rm tr} \frac{1}{- \Box  + \chi},\\
\chi &=&
\frac{\partial V(\eta)}{\partial \eta},
\eea
where $\bar{\eta}$ denotes $\bar{\phi}^2/2N$.
Then we change the variable $\bar{\eta}$ to $\bar{\chi}$
through the Legendre transformation:
\bea
\bar{\chi} &=& \frac{\partial V(\bar{\eta};t)}{\partial \bar{\eta}}, \\
U(\bar{\chi};t) &=& - \bar{\chi} \bar{\eta} + V(\bar{\eta};t).
\eea
Therefore the WH equation for $U(\bar{\chi};t)$ is simply given by
\be
\frac{\partial U}{\partial t}
=DU-2\bar{\chi} U_{\bar{\chi}}+\frac{A_D}{2}\ln (1+\bar{\chi}).
\ee
It is readily seen that this equation is indeed identical to
Eq.~(\ref{largeNLPAWH}) owing to the saddle point equation.

This form of the ERG equation, in turn, is exactly solved by expanding $U$
into an ordinary Taylor series of $\bar{\chi}$ as
\be
U(\bar{\chi};t)=
a_0(t)+a_1(t)\bar{\chi} + \frac{1}{2}a_2(t)\bar{\chi}^2 +\cdots,
\ee
where $a_1$ is just the potential minimum parametrized previously by
$\rho_0$. The relevant operator is found to be $\bar{\chi}$ itself,
which has dimension $D-2$ at the fixed point. The irrelevant operators
are also simply given by $\bar{\chi}^2, \bar{\chi}^3, \cdots$. 
Thus we can reformulate the large $N$ model in terms of the purely
(ir)relevant operators by introducing a new variable, which is a
composite operator of the original scalar fields. On the renormalized
trajectory, we may ignore these irrelevant operators. Once they are
eliminated, the theory turns out to be identical to the non-linear
$\sigma$ model.

What do these relations found in the large $N$ limit imply for the
finite $N$ cases? It would be natural to suppose from the above
observation that good convergence in Scheme II for a finite $N$ may be
explained similarly. Actually, if we compare the forms of the
eigenvectors of the matrix $\Omega$ in Schemes I and II, then we will
see a clear difference. That is, the eigenvectors are approximated
well by the first several components in Scheme II, while this is not
the case in Scheme I. Thus we may deduce that Scheme II is able to
capture the relevant operator in the small dimensional subspace and,
therefore, to make the truncation dependence diminish rapidly.
In practice, it is not hard to extend the formulation of ERG so as to
incorporate the auxiliary field to finite $N$ cases. The results of
numerical analyses of such ERG equations will be reported elsewhere.
To summarize, the physical reason for the good convergence in the
comoving frame is speculated to be that the leading operator
defined in this scheme covers the relevant operator quite well.

\section{Infrared effective potentials}
It is significant to observe the truncation dependence of not only the
exponents but also other physical quantities in various approximation
schemes. We here discuss the infrared effective potentials for the
scalar theory by using the LPA WH equation (\ref{LPAWH}).
The infrared effective potentials enable us to calculate physical
quantities such as the effective mass and the effective couplings at a
low energy scale. The low energy physics is completely described by
the one dimensional renormalized trajectory of the relevant operator
extending from the non-trivial fixed point on the critical surface.
The renormalized trajectory is divided into two parts in the symmetric
phase and in the symmetry broken phase. We evaluate the infrared
effective potentials by tracing the running coupling constants on the
renormalized trajectory. As the cutoff is lowered, the minimum of the
effective potential in the symmetric phase shrinks, while it grows in
the broken phase.

The renormalized trajectories obtained using different truncation
approximations do not coincide. We should be careful when examining
the renormalized trajectory. We need to impose a common appropriate
renormalization condition to compare the infrared effective potentials
evaluated in the various approximation schemes.
Actually, we evaluate the infrared potentials by employing the point
on the renormalized trajectory satisfying the following
renormalization condition; the gradient at the origin of the potential
is equal to $0.1$ in the symmetric phase and the field value at the
minimum of the potential is equal to $0.1$ in the broken phase.
The effective potentials obtained in Schemes I and II are shown in
Figs.~7 and 8, respectively. Note that the absolute height of the
potential is adjusted so as to vanish at the origin, since it is not
taken into account correctly in the ERG equation.\footnote{
In the analysis in terms of Scheme II, the minimum of the potential
moves from the positive region ($\rho >0$) to the negative region
($\rho <0$) in the symmetric phase. Therefore one should switch the
evaluation of the potential using Scheme II to Scheme I at the point
where the minimum of the potential passes the origin $\rho =0$.
}
\begin{figure}[htb]
\parbox{77mm}{
\epsfxsize=0.45\textwidth
\begin{center}
\leavevmode
\epsfbox{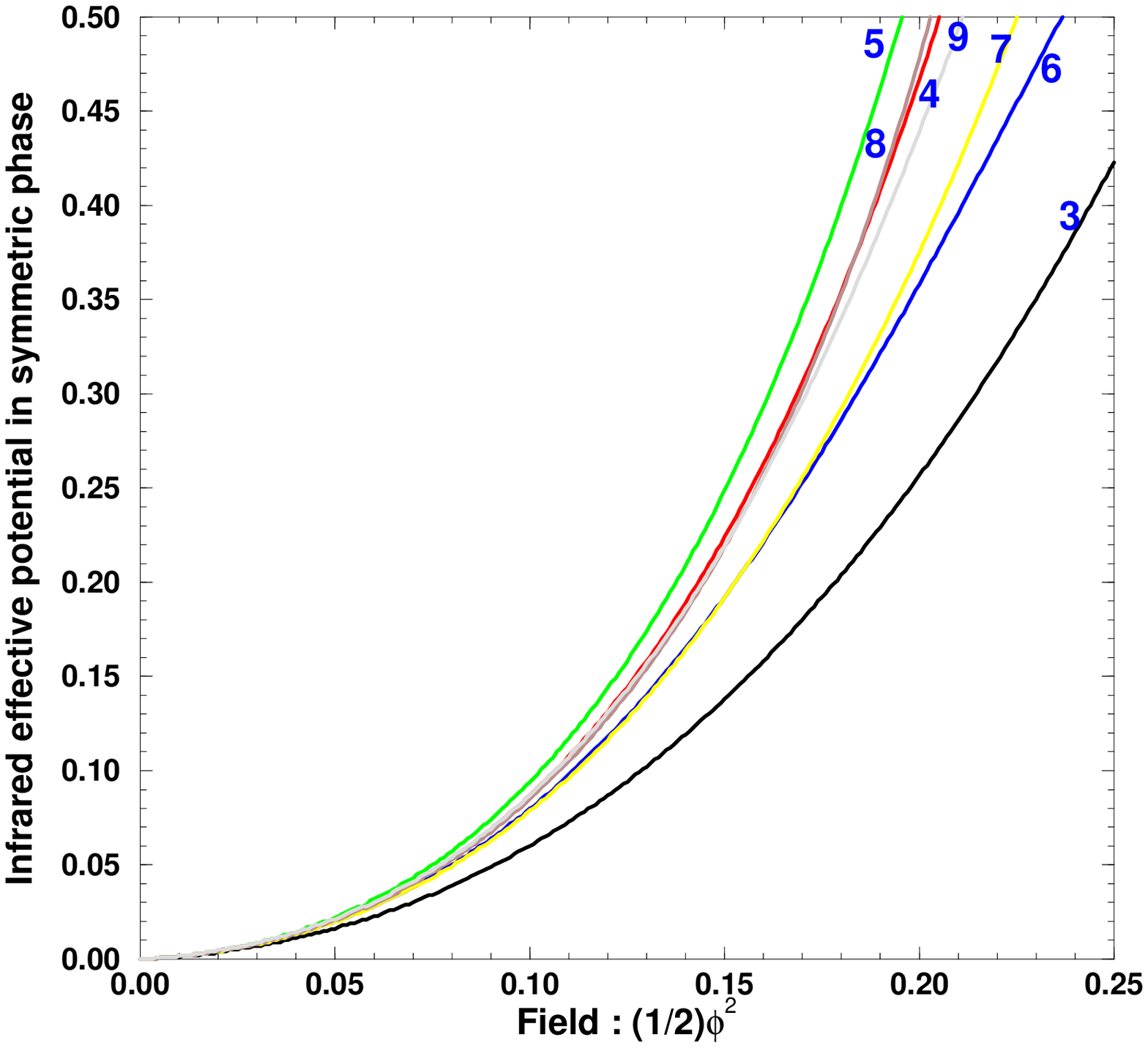}
\end{center}
}
\parbox{77mm}{
\epsfxsize=0.45\textwidth
\begin{center}
\leavevmode
\epsfbox{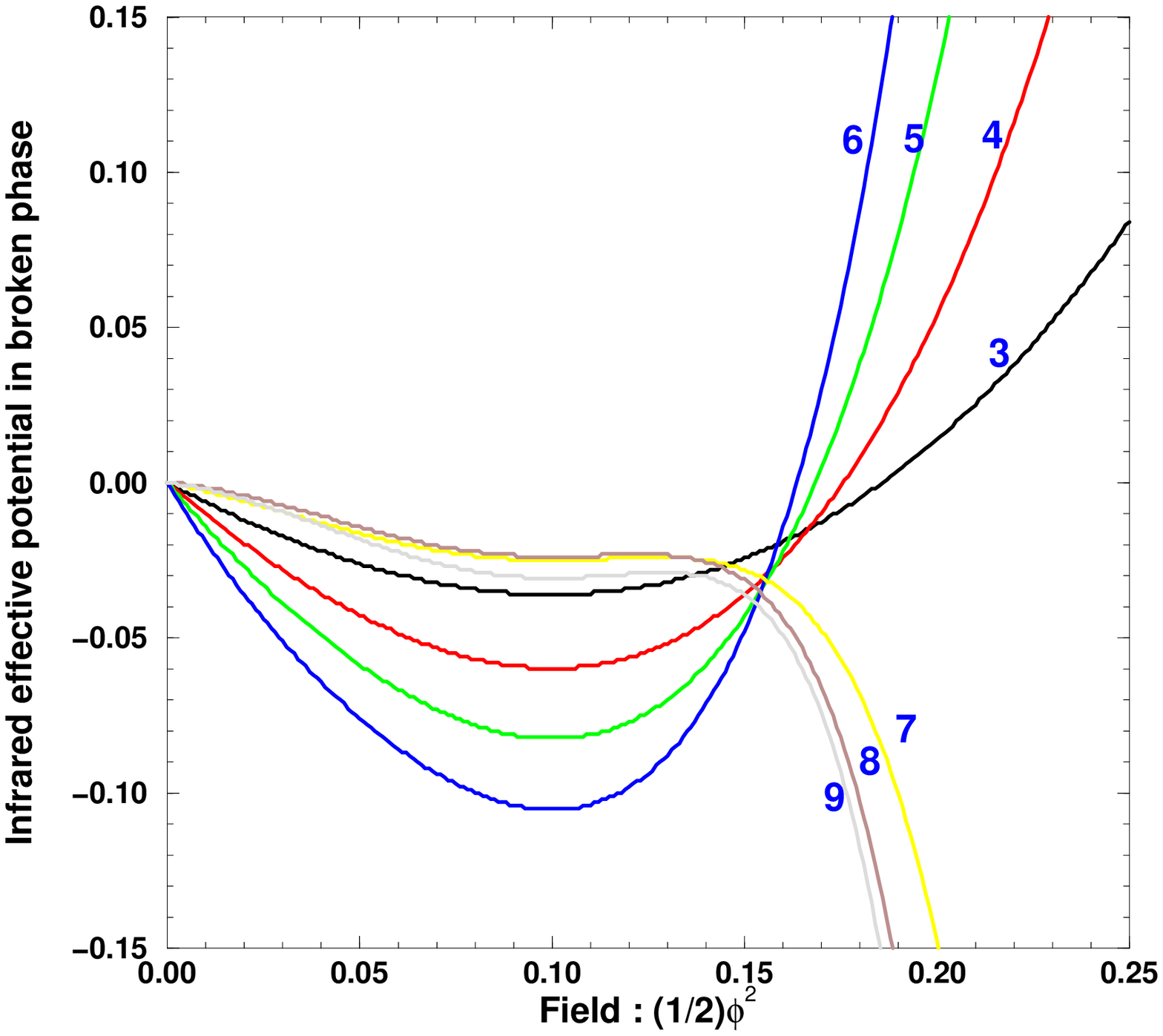}
\end{center}
}
\end{figure}
\vspace{-8mm}
\begin{center}
\parbox{140mm}{
\footnotesize
Fig.~7~~The truncation dependence of the infrared effective potential
in the symmetric and the symmetry broken phases evaluated using
Scheme I
}
\end{center}
\begin{figure}[htb]
\parbox{77mm}{
\epsfxsize=0.45\textwidth
\begin{center}
\leavevmode
\epsfbox{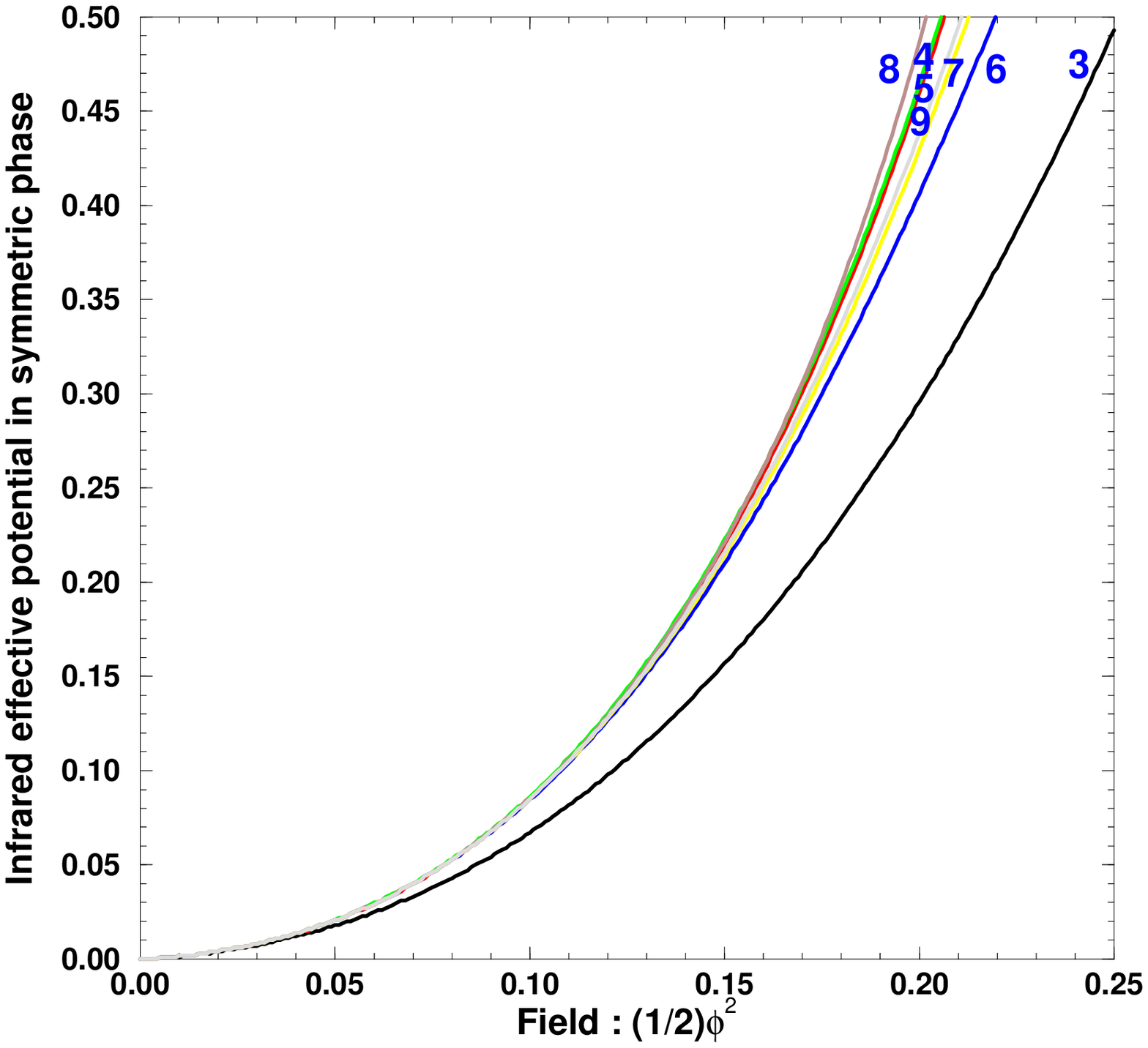}
\end{center}
}
\parbox{77mm}{
\epsfxsize=0.45\textwidth
\begin{center}
\leavevmode
\epsfbox{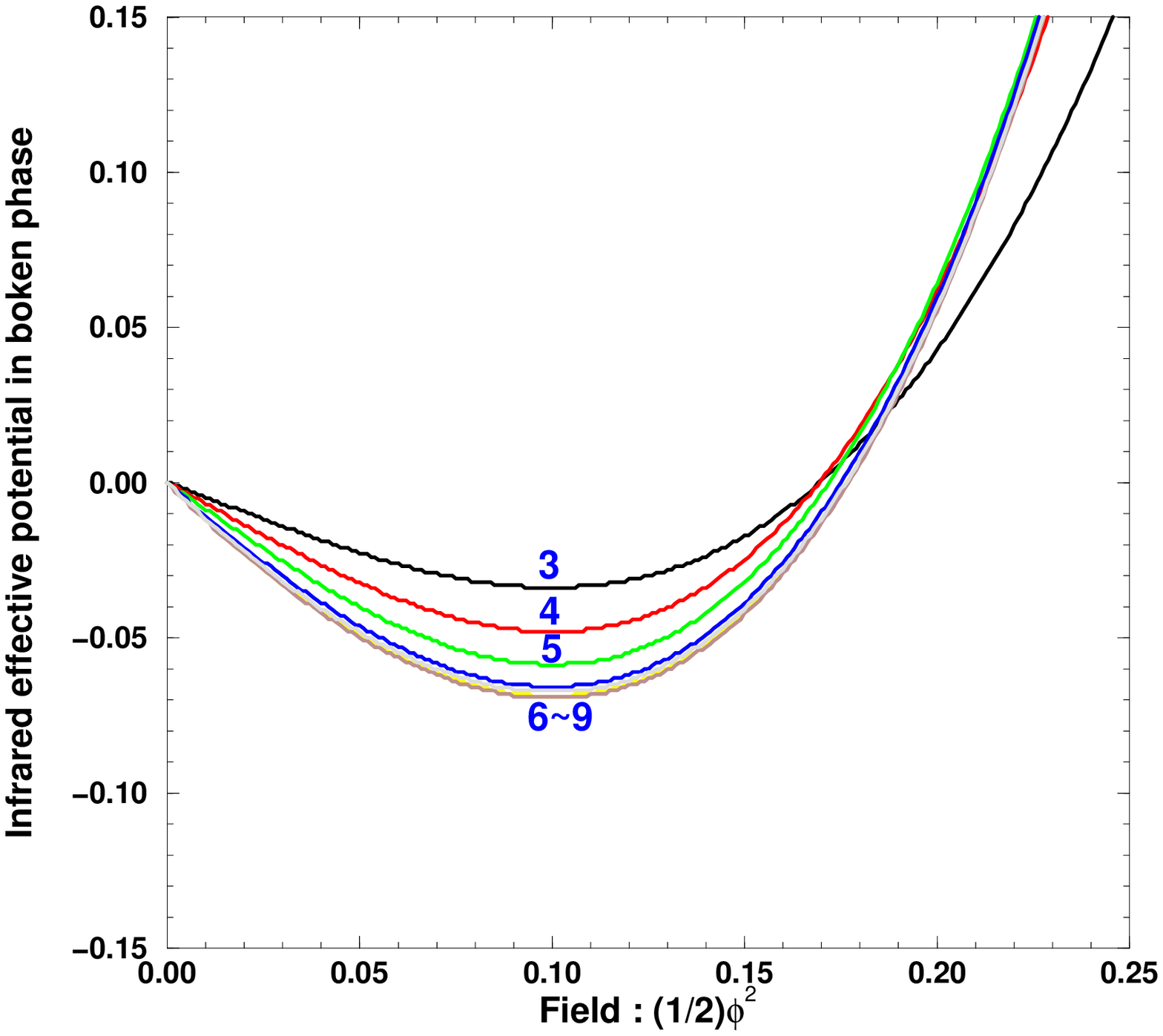}
\end{center}
}
\end{figure}
\vspace{-8mm}
\begin{center}
\parbox{140mm}{
\footnotesize
Fig.~8~~The truncation dependence of the infrared effective potential
in the symmetric and the symmetry broken phases evaluated using
Scheme II
}
\end{center}
\vspace{5mm}

As a result, the truncated approximation in the comoving frame leads
to good convergence for the effective potential in both phases as well
as for the exponent. It is remarkable property of the comoving frame
that it remains so effective after truncation, even away from
criticality, and moreover, that this occurs irrespective of the
phases. This result would imply that the relevant operator ruling the
renormalized trajectory can be approximated well enough in the small
dimensional subspace truncated in the comoving frame.

\section{Summary and discussion}
We considered the convergence properties of physical quantities
evaluated using the ERG in various truncated approximation schemes in
operator expansions. The $Z_2$ symmetric scalar theory in three
dimensions was numerically analyzed, and the approximated solutions
for the exponents and the infrared effective potentials were compared
in the various schemes. In particular we focused on studies of the
difference between the truncation schemes in the expansion at the
field origin (Scheme I) and at the minimum of the effective potential
(Scheme II).

It was found that Scheme II displays a remarkably strong convergence
property with respect to the order of truncation as far as the
quantities we examined are concerned; the leading exponent, the
anomalous dimension and the infrared effective potentials. Although it
is seen that the exponent obtained in Scheme II also ceases to
converge eventually at a certain large order, the width of fluctuation
is very small, and we can obtain the value with great accuracy.

Indeed we may examine the partial differential equations derived in
the derivative expansion scheme directly for such a simple model.
However, such analyses would become difficult for more complicated
systems, e.g., triviality mass bound for the Higgs particle in the
standard model, non-perturbative analysis of dynamical chiral symmetry
breaking of strongly coupled
fermions, etc.\cite{nager,chl,chl-massb,scgt96} ~
Therefore we would like to stress here that the operator expansion
scheme is desired if it gives converging value effectively enough.
Actually, Scheme II is found to satisfy such a practical demand for
scalar theories. It will be necessary to examine the presence of such
a good approximation scheme for the non-perturbative analysis of
various models. Naturally, it would be desirable to seek general
methods to offer us such effective schemes. Such problems in the ERG
approach deserve further study.

We also discussed the physical reason for this rapid convergence in
the comoving frame by considering $O(N)$ symmetric scalar theories in
the large $N$ limit. It was shown that the exponents are derived
exactly by operator expansion in the comoving frame. Not only the
exponents but also each RG flow of the coupling is exactly derived in
every finite order of truncation. Moreover, all of the (ir)relevant
operators at the non-trivial fixed point have been given exactly and
are found to be highly complicated composite operators in terms of the
original scalar fields. We found that the coupling $\rho_0$, the
potential minimum, defined in the comoving frame corresponds to the
exact relevant operator. If such structure remains in the finite $N$
cases in an approximate sense, it could be regarded as the physical
reason for the good convergence of Scheme II. Indeed, the relevant
operator is found to be described well within the subspace of the
first several operators in Scheme II.

The ERG equation in terms of the composite operators has been
proposed. This is equivalent to the ERG equation for scalar theories
in the large $N$ limit. The reformulation is achieved by introducing a 
composite field to the original theory. Interestingly, this operator
turns out to be the exact ``relevant'' operator after evolution to
infrared. Other irrelevant operators are also simply given by the
products of this composite. Namely, the naive polynomial expansion
leads to the perfect coordinates in turn. Thus this offers an example
in which the good expansion scheme in the operators is revealed
through a proper change of field variables. A numerical study of the
ERG in terms of the composite operators in finite $N$ cases will be
reported elsewhere. Indeed, such a variable change incorporating
composite operators has been found to be significant in the RG
analysis of the dynamical chiral symmetry breaking.\cite{scgt96}

\vspace{8mm}
\noindent
{\Large\bf Acknowledgements}

\vspace{4mm}
\noindent
We would like to thank T. R. Morris for valuable discussions at
YKIS '97. Collaboration in the early stages with K. Sakakibara and
Y. Yoshida is also acknowledged. K-I. A. and H. T. are supported in
part by Grants-in-Aid for Scientific Research (\#08240216 and
\#08640361) from the Ministry of Education, Science and Culture.


\end{document}